\documentclass[fleqn,usenatbib]{mnras}

\usepackage{newtxtext,newtxmath}
\usepackage[T1]{fontenc}

\DeclareRobustCommand{\VAN}[3]{#2}
\let\VANthebibliography\thebibliography
\def\thebibliography{\DeclareRobustCommand{\VAN}[3]{##3}\VANthebibliography}

\usepackage{graphicx}	% Including figure files
\usepackage{amsmath}	% Advanced maths commands
\usepackage{physics}

\newcommand{\gaussian}[2]{\mathcal{N}\left( #1, \ #2 \right)}
\title[Colour variability of SNe]{The colour variability of low-$z$ SNe Ia is entirely explained by dust}

\author[Giunta, Karchev \& Trotta]{
Marco Giunta$^{1}$\thanks{E-mail: mgiunta@sissa.it}
Konstantin Karchev,$^{2,1}$ and 
Roberto Trotta$^{1,3,4,5}$
\\
$^{1}$Theoretical and Scientific Data Science, SISSA, Via Bonomea 265, 34136 Trieste, Italy\\
$^{2}$Institute of Cosmos Sciences, University of Barcelona, Martí i Franquès, 1, 08020 Barcelona, Spain\\
$^{3}$INFN -- National Institute for Nuclear Physics, Via Valerio 2, 34127 Trieste, Italy\\
$^{4}$ICSC -- Centro Nazionale di Ricerca in High Performance Computing, Big Data e Quantum Computing, Via Magnanelli 2, Bologna, Italy\\
$^{5}$Astrophysics Group, Physics Department, Blackett Lab, Imperial College London, Prince Consort Road, London SW7 2AZ, UK
}

\date{Accepted XXX. Received YYY; in original form ZZZ}

\pubyear{\the\year{}}

\begin{document}
\label{firstpage}
\pagerange{\pageref{firstpage}--\pageref{lastpage}}
\maketitle

\begin{abstract}
The relative importance of intrinsic colour variability of supernovae type Ia (SN~Ia) versus dust-induced reddening remains an open question with important ramifications for understanding their environmental dependence, as well as for the validity of the traditionally employed Tripp linear correction for cosmological inference.  

We revisit this question in the light of two low-redshift, homogeneous datasets, the ZTF DR2 and Foundation DR1, which we analyse within the framework of the Bayesian hierarchical model Simple-BayeSN. We demonstrate both with simulation and on real data that traditional colour cuts, which remove highly reddened samples, induce a previously unrecognized selection effect, which may have biased previous conclusions on the origin of SN~Ia colour variability. Once this is accounted for, we are able to explain the entirety of the colour--magnitude correlation as due to dust effects, with no need for an intrinsic colour correlation. This result is robust with respect to a host galaxy mass split and projected distance from the center of the host.

Our findings imply that the traditional linear Tripp correction maintains an empirical validity, even though it should be ascribed to dust rather than intrinsic colour variation. 

\end{abstract}

%\begin{keywords}
%keyword1 -- keyword2 -- keyword3
%\end{keywords}

%%%%%%%%%%%%%%%%%%%%%%%%%%%%%%%%%%%%%%%%%%%%%%%%%%

%%%%%%%%%%%%%%%%% BODY OF PAPER %%%%%%%%%%%%%%%%%%

\section{Introduction}

A central question for supernova type Ia (SN~Ia) standardisation is how to apportion the observed colour--magnitude correlation between two physically distinct contributions: an {\em intrinsic} component, reflecting variations in the explosion physics itself, and an {\em extrinsic} component, from interstellar dust in the host galaxy. Traditional approaches, exemplified by the Tripp formula \citep{Tripp_1998}, regress dust-extinguished apparent magnitudes directly against dust-reddened apparent colours using a single linear slope parameter, $\beta$.
The Tripp correction thus does not distinguish between the two physically distinct sources of the observed ``redder--dimmer'' correlation, making it impossible to disentangle their relative importance. This intrinsic-extrinsic colour decomposition is closely entangled with several broader questions in SN~Ia population analysis -- most prominently, the origin of the host-galaxy ``mass-step'' \citep{Kelly_2010, Sullivan_2010, Lampeitl_2010} and the possible variations of dust properties across different galactic environments. These have been the focus of substantial recent work, including the dust-driven mass-step model of \cite{Brout_2021} and \cite{Popovic_2023}, the hierarchical Bayesian analyses of \cite{Grayling_2024}, \cite{Grayling_2025}, and \cite{Ginolin_2026_bayesn}.

\Citet{Mandel_2017} (henceforth M17), following earlier works like \cite{Jha_2007, Mandel_2009, Mandel_2011}, frame the question in the context of Bayesian hierarchical inference and introduce Simple-BayeSN to model SN summary statistics derived with the SALT light-curve fitter \citep{Guy_2007,Guy_2010,Betoule_2014,Taylor_2021,Kenworthy_2021,Taylor_2023}. Simple-BayeSN supposes a Gaussian population of intrinsic colours and an exponential distribution of host-dust reddening. Each of these affects the observed peak brightness: the former through an intrinsic colour--magnitude correlation parameter ($\beta_{\rm int}$) and the latter through the total-to-selective extinction coefficient $R_B$.

Analysing a low-redshift compilation of SN~Ia, M17 inferred a balanced decomposition between intrinsic and extrinsic effects, with $\sigma_{\rm c, int} = 0.067 \pm 0.009$ comparable to $\tau = 0.068 \pm 0.012$. A consequence of this balanced regime is that the joint distribution of dust-extinguished colour and magnitude exhibits a curved, ``banana-like'' shape, transitioning from slope $\beta_{\rm int}= 2.328 \pm 0.262$ in the intrinsic-scatter dominated blue tail to a slope $R_B = 3.758 \pm 0.349$ in the dust-dominated red tail. M17 argued that adopting the single linear Tripp slope $\beta$ therefore induces systematic distance biases in the colour tails, with potentially important consequences for cosmological fits.

This picture was challenged by \cite{Brout_2021} (BS21) and \cite{Popovic_2023}, who modelled the SN~Ia colour distribution on larger compilations using a  two-component framework -- based on the same Gaussian--exponential convolution for colour, but allowing $\beta_{\rm int}$ and $R_B$ to be defined on a per-SN basis. Their analyses recover smaller $\sigma_{\rm c, int}\approx 0.04$ and larger $\tau\approx 0.15$ than M17, so that the dust component plays a more prominent role, but still find a substantial mean intrinsic colour--magnitude correlation $\bar{\beta}_{\rm int} \approx 2$.
A limitation of BS21 and \cite{Popovic_2023} is that they combine SALT-based light curve summaries with hand-tuned forward simulations of selection effects via SNANA \citep{Kessler_2009}, rather than performing fully self-consistent hierarchical inference including selection effects. Establishing the impact of this methodological choice on the recovered population parameters is therefore non-trivial, and exact comparison with M17 impossible.

More recently, a hint that the M17 balanced regime may not be representative of the underlying SN~Ia population emerged from \cite{Ginolin_2025_colour}, who analysed the ZTF SN~Ia DR2 sample \citep{Rigault_2025}. Fitting the same Gaussian--exponential convolution to the empirical colour distribution alone -- without using magnitude information -- they recover $\sigma_{\rm c, int} = 0.030 \pm 0.005$ and $\tau = 0.155 \pm 0.007$, firmly in the regime $\tau \gg \sigma_{\rm c, int}$ rather than the balanced M17 one. They also report $\Delta \beta = -0.71 \pm 0.26$ when fitting separate linear relations to blue and red subsamples (respectively, negative and positive observed colours) -- a modest $2.7\sigma$ significance, indicating a slope change that is much smaller that what would be expected under the M17 balanced regime.

The main reason for this ``surprising'' finding -- as we will demonstrate in this work -- is that in their analysis, \citet{Ginolin_2025_colour} retain heavily reddened SNe with observed colours up to $\hat{c} = 0.8$, whereas historically a conservative cut $\abs{\hat c} \leq 0.3$ has been employed, to avoid extrapolating outside the training regime of the SALT model -- as indeed, the reddest SNe were excluded from the training samples of all major SALT2 \citep{Guy_2007, Betoule_2014, Taylor_2021} and SALT3 \citep{Kenworthy_2021, Taylor_2023} iterations. However, \cite{Rose_2022} argued that SALT2 is actually capable of fitting highly-reddened SNe without significant systematic issues, and recommended against their blanket exclusion -- a recommendation the ZTF collaboration followed in retaining the full colour range in their data release.

While a colour cut is an explicit, analysis-stage truncation of the data, a similar effect can be induced by implicit observational selection effects. For example, a flux limit in the detection of SNe can deplete the red tail of the colour distribution due to the correlation with magnitude: an argument that is independent of the specific light-curve model employed for fitting. This point was emphasised by \cite{Jha_2007}, who also noted that supernova samples assembled from heterogeneous searches without proper modelling of the respective selection functions may be similarly biased against highly extinguished events.

\paragraph*{In this work,} we perform a full hierarchical Bayesian analysis of the ZTF SN~Ia DR2 sample \citep{Rigault_2025}, a homogeneous, volume-limited, low-redshift dataset of unprecedented size. We investigate the relative importance of intrinsic colour variations and dust-induced reddening and -- by retaining the full red tail -- examine the effect that unaccounted-for colour-related selection effects have on the conclusions. We show that in the absence of the traditional colour cut -- or, if applied, when it is correctly accounted for--, the observed magnitude--colour relation is entirely explainable in terms of the effect of dust, with no need for additional intrinsic colour variability.

The remainder of this paper is organised as follows.
Section~\ref{sec:methods} describes the ZTF and Foundation samples employed in this work, the Simple-BayeSN model, our Gibbs and \texttt{emcee} samplers, and our treatment of colour cuts selection through likelihood renormalisation.
Section~\ref{sec:results} presents the population-level inference on ZTF and a comparison with Foundation, which disentangles quality cuts from survey-driven selection effects. In section~\ref{sec:discussion}, we discuss the implications for the Tripp standardisation framework, insights into relations with the host environment, and put our results in the context of previous work.
We conclude in Section~\ref{sec:conclusions}.

\section{Methodology}\label{sec:methods}

\subsection{Data sets}\label{subsec:datasets}

\begin{figure*}
    \centering
    \includegraphics[width=\linewidth]{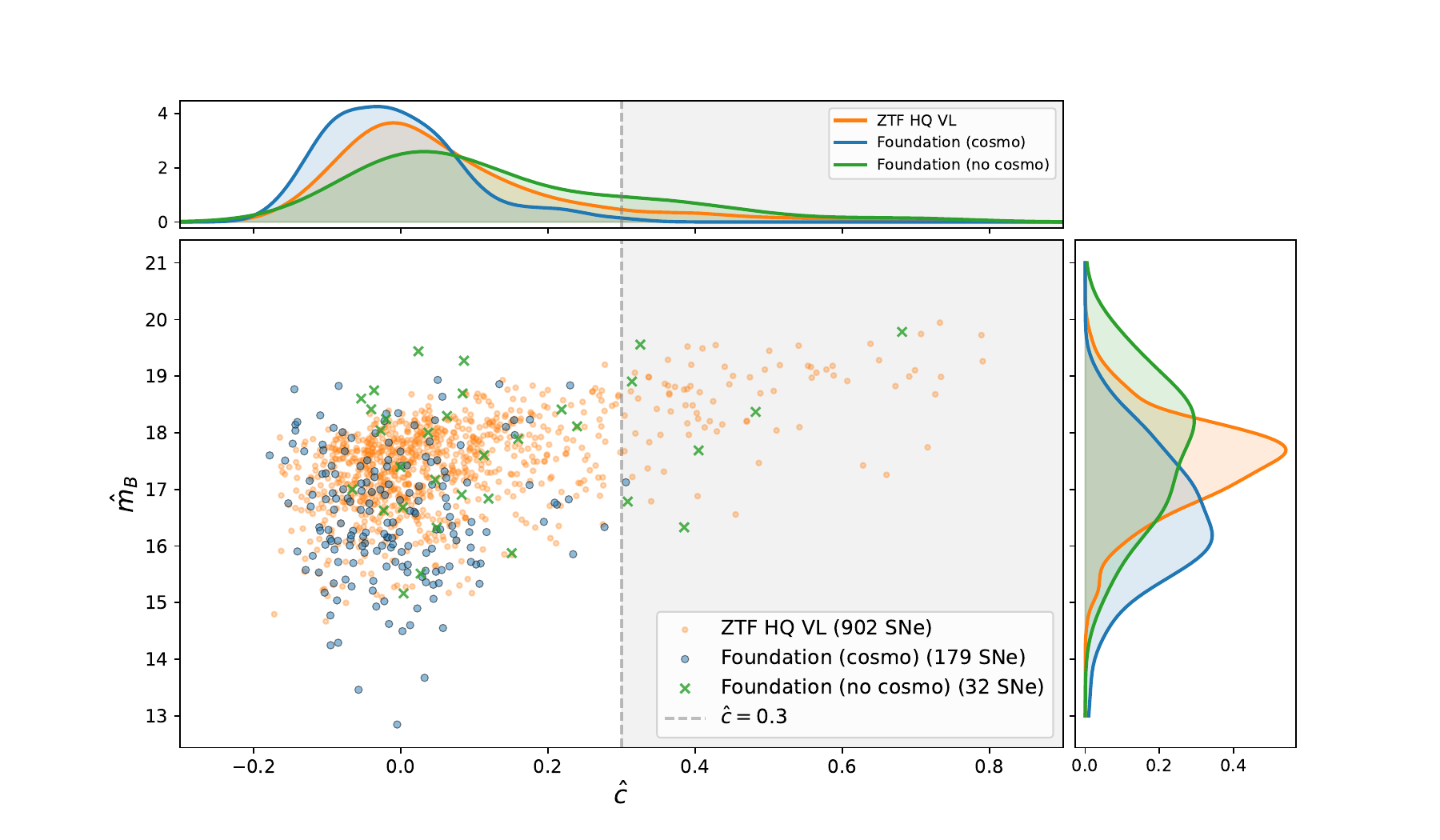}
    \caption{Observed apparent magnitude and colour for the ZTF~HQ~VL, Foundation DR1 ``cosmo'' and ``no cosmo'' samples, with Gaussian KDE marginals. The shaded region marks colours removed by a $\hat{c}\leq 0.3$ quality cut. The magnitude marginals reveal differences between the two surveys that go beyond a simple flux-limit offset, with Foundation containing a population of apparently brighter SNe absent from ZTF~HQ~VL. }
    \label{fig:ztf_foundation_mb_c}
\end{figure*}

\subsubsection{Zwicky Transient Facility DR2}
We analyse the ZTF SN~Ia DR2 sample presented in \cite{Rigault_2025}, the largest homogeneous, publicly available collection of spectroscopically classified low-redshift SNe Ia to date.
We restrict the sample by applying three sets of cuts, described below.

\paragraph*{Light curve coverage and fit quality.}
Following \cite{Ginolin_2025_colour} and \cite{Rigault_2025}, we select supernovae with sufficient light curve coverage and fit quality passing the \texttt{lccoverage} and \texttt{fitquality} requirements defined in the data release. 
The former requires $\geq 7$ detections at $\geq 5\sigma$ from 10 days before to 40 days after peak brightness, including at least two pre-peak and two post-peak detections, distributed across at least two photometric bands.
The latter requires SALT2 fits satisfying $\hat{x}_1\in [-3, 3]$, $\hat{c}\in [-0.2, 0.8]$, $\sigma_{\hat{x}_1}<1$, $\sigma_{\hat{c}}<0.1$, $\sigma_{\hat{t}_0}<1$, and $\texttt{fitprob}\geq 10^{-7}$. As described in \cite{Rigault_2025}, these are derived using SALT2 \citep{Guy_2007, Guy_2010} in its version 4 \citep{Betoule_2014}, retrained by \cite{Taylor_2021} and available via \texttt{sncosmo} \citep{Barbary_2025} as \texttt{SALT2-T21}.

\paragraph*{Volume limited cut.}
We further restrict to $z \leq 0.06$, the redshift below which the ZTF sample is approximately volume-limited and therefore free of Malmquist bias \citep{Amenouche_2025}. 

\paragraph*{Peculiar SN subtypes.}
Following \cite{Ginolin_2026_bayesn}, we remove SN subtypes other than 91T-like and 99aa-like events. In addition, we discard the 53 SNe (5.4\% of the sample) with no assigned subtype classification. As discussed in \cite{Dimitriadis_2025}, most unclassified SNe lack a definite subtype either because their spectra were contaminated by the proximity of the host-galaxy core -- masking the spectral lines required to distinguish peculiar subclasses -- or because their spectra were taken too late after peak, when normal and peculiar SNe Ia become spectroscopically more similar. 
We therefore expect this subset to be a mix of normal and peculiar events, and discard it for simplicity\footnote{Re-including these SNe shifts the posteriors slightly; for instance, the $\beta_{\rm int}$ distribution moves to lower values by $\sim 0.65\sigma$. However, the main conclusions of this work are robust with respect to the removal of these SNe.}.

Together, the three sets of cuts described above define the ZTF~HQ~VL (high quality, volume limited) sample of 902 SNe used throughout this work. The difference with respect to the 938 SNe of \citep{Ginolin_2025_colour} reflects both the slightly more restrictive subtype classification adopted here and the fact that, unlike \cite{Ginolin_2025_colour} and \citep{Ginolin_2026_bayesn}, we retain SNe with missing host-galaxy measurements, since host properties are only needed for the split-sample analysis of Sec.~\ref{subsec:discussion}, not the main population-level inference.

Compared to the heterogeneous low-redshift compilations used in earlier analyses, the resulting sample combines three properties of particular relevance for the present work: homogeneity, near-completeness within the volume-limited regime, and retention of the red colour tail up to $\hat{c}=0.8$, as recommended by \cite{Rose_2022}.
The last of these is especially important for separating intrinsic colour variation from dust extinction, as the red tail is dominated by the exponential dust component and therefore carries most of the information needed to constrain $\tau$ independently of the intrinsic colour spread.

\subsubsection{Foundation DR1}\label{sec:data-foundation}
The Foundation Supernova Survey data \citep{Foley_2018, Jones_2019} shares similar motivations as ZTF: replacing older heterogeneous low-redshift compilations with a large, homogeneous, well-calibrated sample, obtained using the Pan-STARRS1 (PS1) telescope. These properties make the Foundation DR1 sample a natural point of comparison for the ZTF dataset.

The Foundation DR1 sample contains 225 SN~Ia light curves, of which 180 pass the cosmological quality cuts defined in \cite{Foley_2018} (the so-called ``cosmo sample''). For the remaining 45 objects, official SALT2 fits were not published, so we computed them using \texttt{sncosmo}, configured to match the SNANA setup of the Foundation release as closely as possible: we used the same \texttt{SALT2.JLA-B14} model, the included Foundation PS1 photometry, and the same zero-point system and dust law.
We then removed the 12 non-normal SNe Ia as classified in \cite{Foley_2018} and applied the ZTF cuts $-0.2\leq \hat{c}\leq 0.8, \ -3\leq\hat{x}\leq 3$, discarding one further object from the ``no cosmo'' sample (abnormally blue) and one from the ``cosmo'' sample (having $\hat{c}<-0.2$, within the range retained by the original Foundation cuts but outside the ZTF lower bound). This leaves 32 ``no cosmo'' and 179 ``cosmo'' objects, for a total Foundation sample of 211 SNe used in this work.

Figure~\ref{fig:ztf_foundation_mb_c} shows the distributions of observed magnitudes and colours for ZTF HQ VL and Foundation (both the full set and the ``cosmo'' subset). While including ``no cosmo'' objects (green crosses) does extends the Foundation data to redder colours, with a distribution broadly similar to that of ZTF, a domain shift between the two surveys remains, with brighter SNe dominating in Foundation. This can be explained by Foundation's observational selection function and means that accounting solely for the colour cut (or undoing it) can only partially address the differences, as we will show in section~\ref{sec:results}.

\subsection{The Simple-BayeSN model}\label{subsec:simplebayesn}
We outline below the core concepts and equations of the Simple-BayeSN model relevant to our analysis; for a complete description of the model, we refer the reader to M17.

\emph{Simple-BayeSN} is a Bayesian hierarchical model which introduces explicit modelling of the simultaneous extinction and reddening due to dust in the host galaxy: the observed colour distribution is modelled as a convolution of an intrinsic Gaussian component and an extrinsic exponential one, and both are correlated with the peak brightness, albeit with two separate coefficients.

\subsubsection*{Intrinsic SN~Ia population}
For each supernova $s$, the latent intrinsic absolute magnitude $M_s^{\rm int}$ depends on the latent stretch $x_s$ and intrinsic colour $c_s^{\rm int}$ via\footnote{Note that the sign convention for $\alpha$ in Simple-BayeSN is different from most other works.}:
\begin{equation}
    M_s^{\rm int} \sim \gaussian{M_0^{\rm int} + \alpha x_s + \beta_{\rm int}c_s^{\rm int}}{\sigma_{\rm int}^2},
\end{equation}
where $\alpha$ and $\beta_{\rm int}$ are the intrinsic stretch-- and colour--magnitude correlation parameters, $M_0^{\rm int}$ is the standard SN~Ia magnitude, and $\sigma_{\rm int}^2$ is the intrinsic scatter. All are inferred parameters in the model.
The intrinsic colour itself is allowed to depend linearly on stretch:
\begin{equation}
    c_s^{\rm int} \sim \gaussian{c_0^{\rm int} + \alpha_c x_s}{\sigma_{\rm c, int}^2},
\end{equation}
while stretch is defined as independently sampled from a third Gaussian:
\begin{equation}
    x_s \sim \gaussian{x_0}{\sigma_x^2}.
\end{equation}

\subsubsection*{Host galaxy dust}

The latent per-supernova host galaxy dust reddening $E_s \equiv E(B-V)_s\geq 0$ is defined as drawn from an exponential distribution with population mean $\tau = \expval{E_s}$:
\begin{equation}
    E_s \sim P_{\rm exp}(E_s \mid \tau) =
    \begin{cases}
        \frac{1}{\tau}\exp\left(\frac{E_s}{\tau}\right) & E_s \ge 0, \\
        0 & E_s < 0.
    \end{cases}
\end{equation}
This form -- enforcing the physical requirement that dust only dims and reddens and the intuition that most lines of sight encounter little dust -- is one of the most used parametrisations of the host-galaxy dust reddening distribution in hierarchical SN~Ia analyses \citep{Jha_2007, Mandel_2009, Mandel_2011, Mandel_2017, Brout_2021, Popovic_2023}, and we adopt it here both for direct comparison with these prior works and because it admits the closed-form conditional posteriors required for implementing the Simple-BayeSN Gibbs sampler.\footnote{Recent radiative transfer simulations of realistic host-galaxy geometries suggest that two-parameter generalisations such as the Weibull or exponentiated exponential distributions may yield better descriptions of the SN~Ia extinction PDF than the exponential form adopted here \citep{Duarte_2026}; we leave a re--examination of our results under such parametrisations to future work.}

Dust dims and reddens the supernova in a correlated fashion according to:
\begin{align}
    M_s^{\rm ext} &= M_s^{\rm int} + R_B E_s,
    \\ c_s^{\rm app} &= c_s^{\rm int} + E_s,
\end{align}
where $R_B$ is the ratio of $B$-band extinction to $E(B-V)$ reddening. Following M17, we assume for simplicity that $R_B$ is the same for all SNe (similarly to $\beta_{\rm int}$), i.e.\@ that the dust properties in all hosts are the same.

\subsubsection*{The nonlinear dusty colour--magnitude relation}
Crucially, Simple-BayeSN allows $\beta_{\rm int} \neq R_B$, enabling these two slopes to be constrained independently from the data; this has an important geometric consequence in the extinguished colour--magnitude plane. In the absence of dust, SNe Ia scatter around a line of slope $\beta_{\rm int}$.
Host galaxy dust then displaces each supernova by a random exponentially-distributed amount along both colour and magnitude axes with a slope $R_B$; the more a supernova's light is absorbed by dust, the stronger its colour shifts towards larger values -- whereas the supernovae left on the blue side are those which were least affected by dust-induced colour displacement (see Figs.~3 and 6 in M17). 

The overall effect is a curved, banana-like joint distribution of $(c_s^{\rm app}, M_s^{\rm ext})$. In the blue tail, where intrinsic colour scatter $\sigma_{c,{\rm int}}$ dominates over dust reddening, the mean trend follows the slope $\beta_{\rm int}$; in the red tail, where large dust extinction values dominate the apparent colour, the slope tends towards $R_B$. The degree of curvature -- hence the bias incurred by a single linear fit, {\em à la} Tripp -- is controlled by the relative sizes of $\sigma_{c,{\rm int}}$ and $\tau$.

\subsubsection*{Redshift, distance modulus, apparent magnitude}

A SALT fit estimates the SN's apparent rest-frame-$B$-band magnitude:
\begin{equation}
    m_s^B = M_s^{\rm ext} + \mu(z_s^{\rm c}; \Omega),
\end{equation}
where $z_s^{\rm c}$ is the SN's (unknown) cosmological redshift, $\Omega$ are the parameters of the cosmological model, and $\mu$ is the distance modulus, which is a deterministic function of the latter two. Simple-BayeSN assumes a Gaussian redshift measurement $\hat{z}_s \pm \sigma_{z_c,s}$ of $z_c$, with uncertainty coming from both the measurement imperfection ($\sigma_{\hat{z},s}$, reported in the data release) and the correction for peculiar velocities ($\sigma_{\rm pec} = 300 \ \rm km \ s^{-1}$):
\begin{equation}
    \sigma_{z_c,s}^2 = \sigma_{\hat{z},s}^2 + \sigma_{\rm pec}^2 / c^2,
\end{equation}
where $c$ is the speed of light. This is then propagated through the distance modulus function under the assumption of linear cosmological expansion (appropriate for low redshifts $\lesssim 0.1$):
\begin{equation}
    \sigma_{\mu|\hat{z},s}^2 = \left(\frac{5}{\hat{z}_s\ln 10}\right)^2 
    \left[\sigma_{\hat{z}}^2 + \frac{\sigma_{\rm pec}^2}{c^2}\right],
\end{equation}
and the distance modulus is promoted to a latent variable, to be inferred for each SN:
\begin{equation}
    \mu_s \sim \mathcal{N}\!\left(
    \mu(\hat{z}_s;\Omega), \sigma_{\mu|\hat{z},s}^2\right).
\end{equation}
When spectroscopic redshifts are available ($\sigma_{\hat{z}, s} \sim 10^{-5}$, relevant for both the ZTF and Foundation samples), this approximation is sufficient. It is also necessary to ensure tractability of the Gibbs sampler we use. However, with less precise photometric estimates, linear error propagation can introduce a major systematic bias \citep{Karchev_2022}.

In this work, which only examines the SN~Ia population, following \citet{Ginolin_2025_colour}, we fix the cosmological model to that inferred by \cite{Planck_2020} and compute $\mu(\hat{z}_s;\Omega)$ using the corresponding routine in \texttt{astropy} \citep{astropy:2013, astropy:2018, astropy:2022}.

\subsubsection*{Observables and measurement errors}

The three light-curve summaries -- peak $B$-band apparent magnitude $\hat{m}_s^{B}$, apparent colour $\hat{c}_s^{\rm app}$, and stretch $\hat{x}_s$, as obtained by fitting the SALT2 light-curve model -- are related to the latent parameters via:
\begin{equation}
    \hat{\vb{d}}_s \sim \gaussian{\hat{\vb{d}}_s \mid \vb*{\varphi}_s}{\hat{\vb{W}}_s}
\end{equation}
where $\hat{\vb{d}}_s \equiv (\hat{m}_s^B, \hat{c}_s^{\rm app}, \hat{x}_s)$, $\vb*{\varphi}_s \equiv (M_s^{\rm ext} + \mu_s, c_s^{\rm app}, x_s)$, and the covariance matrix $\hat{\vb{W}}_s$ captures the correlated uncertainties in the best-fit parameters.

\subsection{High-dimensional inference using Gibbs sampling}

Our inference goals are twofold: to derive the posterior probabilities of global/population parameters of the Simple-BayeSN model, after setting priors as in table~\ref{tab:priors}; and to examine and visualise the recovered values for all latent parameters. To achieve these, we implement a Gibbs sampling scheme analogous to that of M17, which can represent the full $5N_{\text{SNe}}+11$-dimensional posterior distribution.
As detailed in M17, all required conditional distributions are analytical and can be efficiently sampled, and by implementing vectorised handling of the latent parameters, we can draw $10^5$ posterior samples for the ZTF HQ VL data set ($\approx 1000$ SNe) in two minutes on an AMD Ryzen AI 9 365 laptop CPU.

We run four independent chains, which converge effectively at initialisation, and we discard only the first $10^3$ samples as a nominal burn-in; the Gelman--Rubin statistic (see e.g.~\cite{Trotta_2008}) is $\hat{R}\leq 1.01$ for all parameters across all runs. We also test the Gibbs sampler on simulations drawn from the Simple-BayeSN forward model with known input parameters, recovering them well within the credible intervals -- an example is shown in Sec.~\ref{subsec:cuts_induced_biases}.

\begin{table}
    \centering
    \begin{tabular}{c|c}
        \hline
        Parameter & Prior\\
        \hline
        $\tau$ & $\Gamma^{-1}(0.003, 0.003)$\\
        $R_B$ & $U(2, 4)$\\
        $x_0$ & $U(-1, 0)$\\
        $\sigma_x^2$ & $\Gamma^{-1}(0.003, 0.003)$\\
        $c_0^{\rm int}$ & $U(-1, 0)$\\
        $\alpha_c^{\rm int}$ & $U(-1, 0)$\\
        $\sigma_{\rm c, int}^2$ & $\Gamma^{-1}(0.003, 0.003)$\\
        $M_0^{\rm int}$ & $U(-21, -18)$\\
        $\alpha$ & $U(-0.2, -0.1)$\\
        $\beta_{\rm int}$ & $U(-3, 3)$\\
        $\sigma_{\rm int}^2$ & $\Gamma^{-1}(0.003, 0.003)$\\
    \end{tabular}
    \caption{Priors for global parameters. Wide inverse-Gamma priors on variance parameters with both shape and scale set to $0.003$ are employed to weakly enforce a positivity constraint.}
    \label{tab:priors}
\end{table}

\subsection{Selection effects in SN~Ia colour}\label{subsec:selection_effects_color}

Correct inference for any population-level distribution depends on whether the observed sample is representative of the underlying population. In the context of Simple-BayeSN, the $\abs{\hat c}\leq 0.3$ quality cut commonly applied to SALT2 fits truncates the red tail of the apparent colour distribution, which carries the most information about dust reddening. If this is not explicitly taken into account during the analysis, the inferred $\tau$ will be biased towards smaller values.
Moreover, the two dimmer--redder effects will remain intertwined, possibly biasing the inferred $\beta_{\rm int}$ towards $R_B$, as we now demonstrate.
In order to implement the necessary selection effect correction (which would break the Gibbs sampling scheme), we need to marginalise out all per-SN latent variables so that the resulting marginal likelihood is amenable to sampling by Markov Chain Monte Carlo. This proceeds as follows.

The Simple-BayeSN model -- provided the observed redshift $\hat{z}_s$ and SALT2 fit covariance matrix $\hat{\vb{W}}_s$ -- admits a closed-form marginal likelihood 
$\mathcal{L}_{\rm sbsn}(\hat{\vb{d}}_s \mid \hat{z}_s, \hat{\vb{W}}_s, \vb*{\theta})$ for the 11 global/population parameters
\begin{equation}
    \vb*{\theta} = (\tau, R_B, x_0, \sigma_x^2, c_0^{\rm int}, \alpha_c^{\rm int}, \sigma_{\rm c, int}^2, M_0^{\rm int}, \alpha, \beta_{\rm int}, \sigma_{\rm int}^2),
\end{equation}
given the data vector $\hat{\vb{d}}_s$ of supernova $s$. The complete expression, reported in Eq.~37 of M17, is obtained by analytically marginalising over all supernova-specific latent variables $\vb*{\lambda}_s=(M_s^{\rm ext}, c_s^{\rm app}, x_s, E_s, \mu_s)$.

Due to the assumed conditional independence of the SNe, the likelihood given the full data set is then simply
\begin{equation}
    \mathcal{L}_{\rm sbsn}\left(\hat{\vb{d}} \mid \hat{\vb{z}}, \hat{\vb{W}}, \vb*{\theta}\right) = \prod_{s=1}^{N_s} \mathcal{L}_{\rm sbsn}\left(\hat{\vb{d}}_s \mid \hat{z}_s, \hat{\vb{W}}_s, \vb*{\theta}\right),
\end{equation}
where we have denoted $\hat{\vb{z}} \equiv [\hat{z}_s]_{s=1}^{N_s}$, $\hat{\vb{W}} \equiv [\hat{\vb{W}}_s]_{s=1}^{N_s}$, and $\hat{\vb{d}} \equiv [\hat{\vb{d}}_s]_{s=1}^{N_s}$.

This also allows us to validate the marginal posteriors of our Gibbs sampler against the results of low-dimensional affine-invariant Markov chain Monte Carlo (MCMC) sampling, as implemented in the \texttt{emcee} package \citep{emcee}. The two methods produce perfectly consistent posteriors.

\paragraph*{Accounting for colour and stretch cuts.} Discarding data points from a sample of SALT2 fits if the fit colour and stretch lie outside certain ranges can be thought of as applying a ``box'' selection function: 
\begin{multline}
    \label{eq:box_selection_density}
    P(S_s=1 \mid \hat{\vb{d}}_s, \hat{z}_s, \hat{\vb{W}}_s, \vb*{\theta})
    = S_{\rm box}(\hat{\vb{d}}_s)
    \\ = \mathbb{I}(c_{\rm min} < \hat{c}_s < c_{\rm max}) \times \mathbb{I}(x_{\rm min} < \hat{x}_s < x_{\rm max}),
\end{multline}
where $S_s=1$ indicates that supernova $s$ passed the selection. Then, by Bayes' theorem, the likelihood becomes:
\begin{equation}\label{eq:renormalised_likelihood}
    P(\hat{\vb{d}}_s \mid \hat{z}_s, \hat{\vb{W}}_s, \vb*{\theta}, S_s=1) = \frac{P(S_s=1 \mid \hat{\vb{d}}_s, \hat{z}_s, \hat{\vb{W}}_s, \vb*{\theta}) P(\hat{\vb{d}}_s \mid \hat{z}_s, \hat{\vb{W}}_s, \vb*{\theta})}{P(S_s=1 \mid \hat{z}_s, \hat{\vb{W}}_s, \vb*{\theta})}.
\end{equation}

For a selection criterion like Eq.~\ref{eq:box_selection_density}, which does not depend on unknown parameters $\vb*{\theta}$, the selection term in the numerator is, by definition, unity for SNe that end up in the analysed data. The important factor remains the renormalisation in the denominator:
\begin{equation}
    \begin{split}
    P(S_s=1 \mid \hat{z}_s, \hat{\vb{W}}_s, \vb*{\theta}) 
    &= \int P(S_s=1 \mid \hat{\vb{d}}_s, \hat{z}_s, \hat{\vb{W}}_s)\, P(\hat{\vb{d}}_s \mid \hat{z}_s, \hat{\vb{W}}_s, \vb*{\theta})\, \dd{\hat{\vb{d}}_s} \\
    &= \int S_{\rm box}(\hat{\vb{d}}_s)\, P(\hat{\vb{d}}_s \mid \hat{z}_s, \hat{\vb{W}}_s, \vb*{\theta})\, \dd{\hat{\vb{d}}_s} \\
    &= \mathbb{E}_{\hat{\vb{d}}_s \sim P(\cdot \mid \hat{z}_s, \hat{\vb{W}}_s, \vb*{\theta})}
       \left[ S_{\rm box}(\hat{\vb{d}}_s) \right],
    \end{split}
\end{equation}
which represents the average probability of selecting the SN, given certain global/population parameters. It can be approximated using a simple Monte Carlo scheme:
\begin{enumerate}
    \item Draw $N_{\rm sim}$ simulated observations for supernova $s$ from the Simple-BayeSN generative model conditioned on $\vb*{\theta}$ and the per-SN quantities $\hat{z}_s, \hat{\vb{W}}_s$:
    \begin{equation}
        \{\hat{\vb{d}}_s^{\,(i)}\}_{i=1}^{N_{\rm sim}} \sim P\left(\cdot \mid \hat{z}_s, \hat{\vb{W}}_s, \vb*{\theta}\right).
    \end{equation}
    \item Estimate the selection probability as the fraction of simulated samples that pass the selection cuts:
    \begin{equation}
        P\left(S_s=1 \mid \hat{z}_s, \hat{\vb{W}}_s,\vb*{\theta}\right) \approx \frac{1}{N_{\rm sim}}\sum_{i=1}^{N_{\rm sim}} S_{\rm box}\left(\hat{\vb{d}}_s^{\,(i)}\right).
    \end{equation}
\end{enumerate}
Since each supernova's selection probability is estimated independently using its own $(\hat{z}_s, \hat{\vb{W}}_s)$, this algorithm can be efficiently parallelised on a graphics processing unit (GPU).

The renormalised likelihood for the full data set is then:
\begin{equation}
    \mathcal{L}_{\rm sel}\left(\hat{\vb{d}} \mid \vb{S}=\vb{1}, \hat{\vb{z}}, \hat{\vb{W}}, \vb*{\theta}\right) = 
    \prod_{s=1}^{N} \frac{S_{\rm box}(\hat{\vb{d}}_s)\, 
    \mathcal{L}_{\rm sbsn}\left(\hat{\vb{d}}_s \mid \hat{z}_s, \hat{\vb{W}}_s, \vb*{\theta}\right)}
    {P(S_s=1 \mid \hat{z}_s, \hat{\vb{W}}_s, \vb*{\theta})},
\end{equation}
where again $\vb{S}\equiv[S]_{s=1}^{N_s}$, and can be used for MCMC as before. 

\section{Results}\label{sec:results}

First, we demonstrate the impact of colour cuts on posterior population inference, and that our likelihood renormalisation procedure correctly recovers the true parameters in the presence of selection cuts from simulated ZTF-like data.

\subsection{Colour cuts-induced $\tau$ and $\beta_{\rm int}$ biases}\label{subsec:cuts_induced_biases}

\begin{figure*}
    \centering
    \includegraphics[width=1\linewidth]{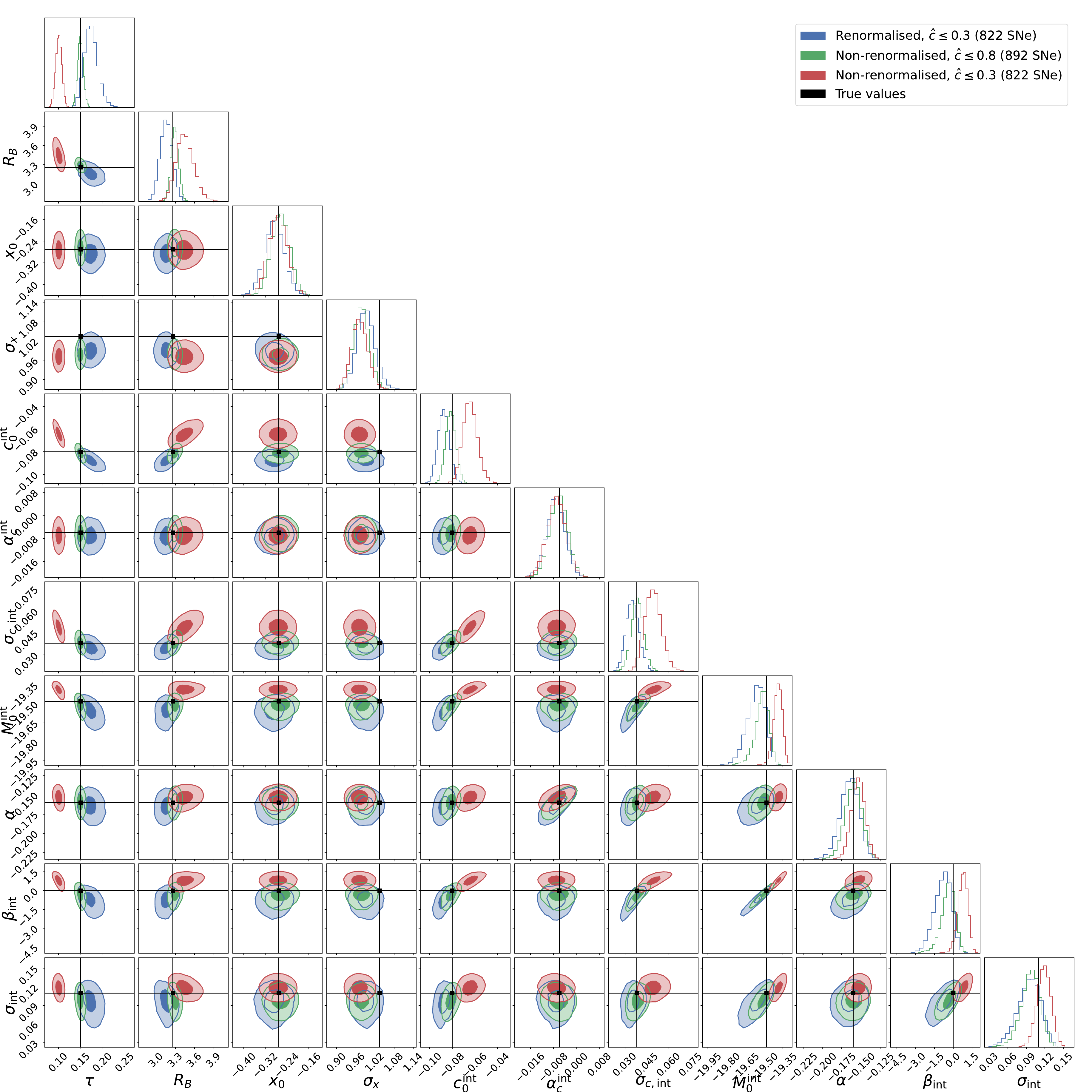}
    \caption{Cornerplot comparison of inferred Simple-BayeSN population parameters from the simulated datasets under different treatments of selection effects. The input population is identical in all cases, while the observed sample is truncated in colour to mimic either the $\abs{\hat{c}_{\rm app}}\leq 0.3$ selection (traditional quality cuts) or $\hat{c}_{\rm app}\in[-0.3, 0.8]$ (ZTF-like cuts). Both samples are cut according to the usual $\abs{\hat{x}}\leq 3$ for simplicity.
    Results are shown for (i) inference on the $\hat{c}\leq 0.3$ dataset with likelihood renormalisation (emcee sampler with MC estimate of the $\prod_i P(S_i=1|\vb*{\theta})$ integral), (ii) the base Gibbs sampler on the $\hat{c}\leq 0.8$ dataset (i.e. without renormalisation), and (iii) the same Gibbs sampler on the $\hat{c}\leq 0.3$ sample.
    Without proper renormalisation to account for the truncated support, the inferred parameters are biased, with $\tau$ underestimated and $\beta_{\rm int}$ correspondingly shifted to larger values incompatible with the truth. Including the renormalisation restores unbiased recovery of the input population parameters.}
    \label{fig:ztf_sim_bias}
\end{figure*}

In figure \ref{fig:ztf_sim_bias}, we show an example of the $\vb*{\theta}$ inference when renormalising the $\mathcal{L}_{\rm sbsn}(\hat{\vb{d}}_i | \vb*{\theta})$ likelihood to account for colour and stretch cuts. Posteriors were obtained from a ZTF-like dataset simulated using the Simple--BayeSN generative process, conditioned on 
\begin{enumerate}
    \item the vectors of redshifts, redshift uncertainties, and SALT2 covariance matrices of the ZTF~HQ~VL sample;
    \item the values of the global parameters shown in table \ref{tab:ztf_sim_truth}.
\end{enumerate}

The values in table \ref{tab:ztf_sim_truth} are chosen to be approximately equal to the ZTF posterior mean estimates obtained from the Gibbs sampler, as shown in Tab.~\ref{tab:ztf_mean_std} -- with the notable exception of $\beta_{\rm int}$, which is set exactly to zero to study what happens when the true colour--magnitude correlation is entirely due to dust.

\begin{table}
    \centering
    \begin{tabular}{cc}
        Parameter & True value  \\\hline
        $\tau$ & 0.15 \\\hline
        $R_B$ & 3.26 \\\hline
        $x_0$ & -0.27 \\\hline
        $\sigma_x$ & 1.03\\\hline
        $c_0^{\rm int}$ & -0.08\\\hline
        $\alpha_c^{\rm int}$ & -0.006\\\hline
        $\sigma_{\rm c, int}$ & 0.038\\\hline
        $M_0^{\rm int}$ & -19.48 \\\hline
        $\alpha$ & -0.16 \\\hline
        $\beta_{\rm int}$ & 0 \\\hline
        $\sigma_{\rm int}$ & 0.11\\\hline
    \end{tabular}
    \caption{Table of fiducial values used to generate the simulated sample shown in fig.~\ref{fig:ztf_sim_bias}, with Simple--BayeSN in the forward fashion used as the generative model.}
    \label{tab:ztf_sim_truth}
\end{table}

Figure \ref{fig:ztf_sim_bias} compares three inference algorithms:
\begin{enumerate}
    \item An \texttt{emcee}--based sampler using the renormalised likelihood \eqref{eq:renormalised_likelihood} and the simulated data \emph{with cuts};
    \item The standard (i.e. not renormalised) Simple--BayeSN Gibbs sampler using the simulated data \emph{without cuts};
    \item The same Gibbs sampler, but given data \emph{with cuts}.
\end{enumerate}

Compared to the reference case (Gibbs sampler, no data cuts), the renormalised sampler is able to recover unbiased estimates of $\tau$ and $\beta_{\rm int}$. Failing to account for the data cuts, instead, biases $\tau$ downwards and $\beta_{\rm int}$ upwards. This is an intuitive result: $\tau$ measures the length of the empirical colour distribution's right tail, and removing highly-reddened SNe depletes precisely the same part of the distribution. As a consequence, the model reassigns part of the overall (positive) colour--magnitude correlation to the intrinsic component, biasing $\beta_{\rm int}$ away from its true value 0. These results help explain why Tab.~1 in M17 estimated $\tau$ to be approximately half compared to our results, and $\beta_{\rm int}$ as a positive number $\sim 9\sigma$ away from 0.

\subsection{Intrinsic colours in ZTF}

We now present findings from real data. Our complete inference results for global/population parameters derived with Simple-BayeSN from the ZTF~HQ~VL sample (with or without the $\hat{c}\leq 0.3$ colour cut applied) are shown in Appendix~\ref{app:cornerplots} and summarised in Tab.~\ref{tab:ztf_mean_std}. 

\begin{figure}
    \centering
    \includegraphics[width=1\linewidth]{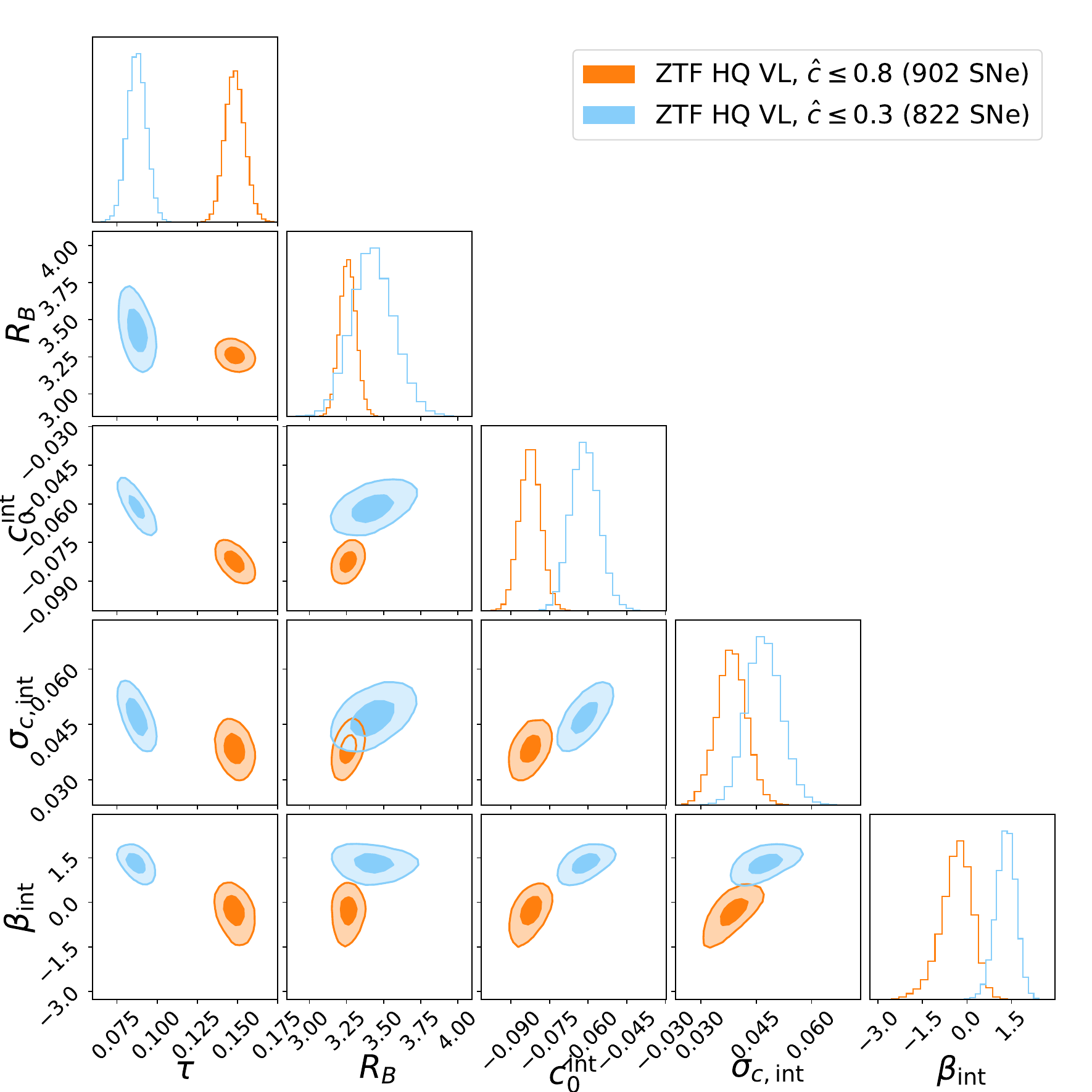}
    \caption{Marginal posteriors for colour-related parameters as fit with the Simple-BayeSN model using the ZTF~HQ~VL sample, with and without the $\hat{c}\leq 0.3$ colour cut.}
    \label{fig:ztf_color_params_cornerplot}
\end{figure}

A corner plot for the colour-related parameters is shown in figure \ref{fig:ztf_color_params_cornerplot}. The inferred values of $c_0^{\rm int}$, $\sigma_{\rm c, int}$, and $\tau$ obtained without the colour cut are in general agreement with \cite{Ginolin_2025_colour} (despite minor differences in the data sets used due to slightly different application of quality cuts, as explained above). By contrast, applying the $\abs{\hat{c}}\leq 0.3$ cut shifts the posteriors markedly: $\tau$ decreases to $0.087 \pm 0.006$ while $\beta_{\rm int}$ increases to $1.309 \pm 0.341$, consistent with the bias discussed in Sec.~\ref{subsec:cuts_induced_biases}.

\begin{table}
    \centering
    \begin{tabular}{c|c|c}
        Parameter & Simple--BayeSN & \cite{Ginolin_2025_colour} \\\hline
        $\tau$ & $0.148 \pm 0.006$ & $0.155 \pm 0.007$ \\\hline
        $R_B$ & $3.260 \pm 0.055$ & -- \\\hline
        $x_0$ & $-0.274 \pm 0.035$ & -- \\\hline
        $\sigma_x$ & $1.032 \pm 0.025$ & -- \\\hline
        $c_0^{\rm{int}}$ & $-0.083 \pm 0.004$ & $-0.085 \pm 0.004$ \\\hline
        $\alpha_c^{\rm{int}}$ & $-0.006 \pm 0.003$ & -- \\\hline
        $\sigma_{c, \rm{int}}$ & $0.038 \pm 0.004$ & $0.030 \pm 0.005$\\\hline
        $M_0^{\rm{int}}$ & $-19.483 \pm 0.053$ & -- \\\hline
        $\alpha$ & $-0.159 \pm 0.011$ & -- \\\hline
        $\beta_{\rm{int}}$ & $-0.360 \pm 0.533$ & -- \\\hline
        $\sigma_{\rm{int}}$ & $0.113 \pm 0.015$ & -- \\\hline
        \end{tabular}
    \caption{Posterior means and standard deviations for all Simple-BayeSN population parameters inferred from the ZTF~HQ~VL sample used in this work (902 SNe). For comparison, we also report the values of $c_0^{\rm int}$, $\sigma_{\rm c, int}$ and $\tau$ from the ``full sample'' entry in Tab.~1 of \protect\cite{Ginolin_2025_colour}, who fit the same Gaussian--exponential convolution to the empirical colour distribution of a similar ZTF subset (938 SNe).}
    \label{tab:ztf_mean_std}
\end{table}

Several results are worth noticing. 
Firstly, we find that $\beta_{\rm{int}}$ is compatible with zero at the $0.67\sigma$ level, in stark contrast with the $\beta_{\rm{int}}=2.328\pm 0.262$ value reported in M17, who used a heterogeneous low-redshift compilation drawn from CfA1--4, CSP, SDSS, SNLS, and Pan-STARSS \citep{Scolnic_2015}, subject to the $\abs{\hat{c}}\leq 0.3$ colour cut.

Our result challenges the conventional SN~Ia wisdom that part of the ``redder--dimmer'' correlation is intrinsic to type Ia supernovae, rather ascribing it entirely to the effect of dust.

Secondly, we find a significantly larger value for the average dust-induced reddening compared to $\tau=0.068\pm 0.012$ in M17, and a tiny value for the the intrinsic colour spread; these results imply, consistently with the previous point, that the colour--magnitude correlation emerging from SALT2 is almost entirely due to dust, and that the intrinsic colour is actually approximately the same across the full SN~Ia population.
These two points are related: a larger $\tau$ apportions a larger part of the observed colour--magnitude correlation to the extrinsic component, leaving less to be explained by the intrinsic slope.
In general, the inclusion of the red colour tail -- previously ignored -- directly shapes the model's decomposition of the two competing effects.

Still, care is needed when interpreting these conclusions. With $\sigma_{\rm c, int} = 0.038 \pm 0.004$, the intrinsic colour component in our analysis is small and nearly constant across the population. Such a small effect size trivially shows little correlation with anything else. In this sense, our $\beta_{\rm int}\approx 0$ result is at least partly a consequence of the dust-dominated regime $\tau \gg \sigma_{\rm c, int}$ that the same data imply: the intrinsic colour simply does not vary enough to carry a strong magnitude correlation.
This consideration is not entirely conclusive, however. \cite{Brout_2021} and \cite{Popovic_2023} recover comparable values of $\sigma_{\rm c, int}$ and $\tau$ on their respective samples, yet still infer $\bar{\beta}_{\rm int}\approx 2$ -- showing that a small intrinsic colour spread is not, by itself, sufficient to force $\beta_{\rm int}$ to zero. Our central claim is to identify the regime $\tau \gg \sigma_{\rm c, int}$ and its consequences for the Tripp standardisation, which are robust to the precise value of $\beta_{\rm int}$.

\begin{figure}
    \centering
    \includegraphics[width=1\linewidth]{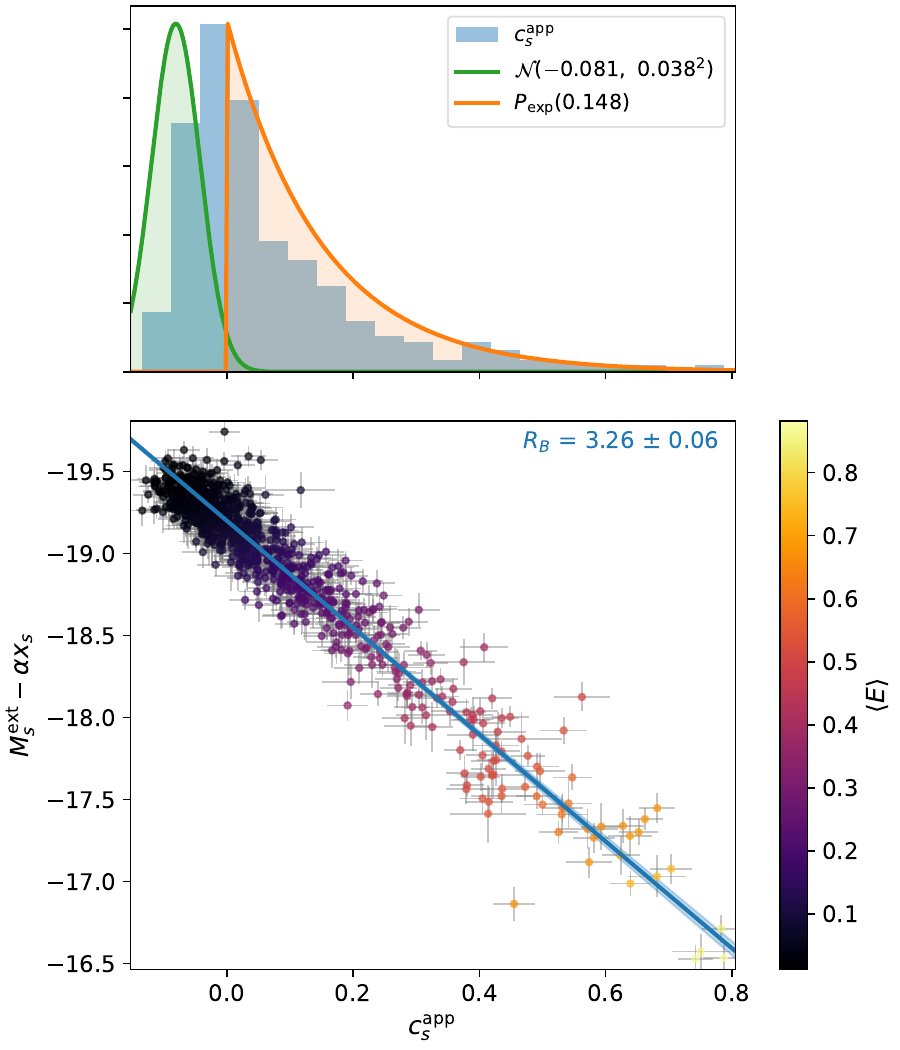}
    \caption{\textit{Top panel:} Distribution of posterior mean latent apparent colours $c_s^{\rm app}$ for the 902 SNe in the ZTF~HQ~VL sample used throughout this work (blue histogram), overlaid with the inferred intrinsic Gaussian component $\mathcal{N}(c_0^{\rm int},\, \sigma_{c,\rm int}^2)$ (green) and the exponential dust component $P_{\rm exp}(\tau)$ (orange), each scaled to the histogram peak for visual comparison.
    The dust component dominates the colour spread, with $\tau\approx 4\sigma_{\rm c, int}$. \\
    \textit{Bottom panel:} Marginal posteriors ($\text{means}\pm\text{st.dev.}$) for stretch-corrected absolute magnitudes and dust-extinguished colours for the 902 SNe in the ZTF~HQ~VL sample. The blue line depicts the marginal posterior for $R_B$, as labelled in the top-right corner, while the colour scale depicts the posterior average for each $E_s$.}
    \label{fig:ext_ztf}
\end{figure}

As discussed in Sec.~\ref{subsec:simplebayesn}, the degree of curvature of the joint colour--magnitude distribution is controlled by the relative sizes of $\tau$ and $\sigma_{\rm c, int}$, with the M17 balanced regime ($\tau = 0.068 \pm 0.012 \approx \sigma_{\rm c, int} = 0.067 \pm 0.009$) producing a pronounced banana visible in their Fig.~6. When instead $\tau\gg\sigma_{\rm c, int}$, the dust component overwhelmingly dominates the colour spread, the intrinsic-scatter-dominated regime is confined to a narrow blue region, and the distribution appears near linear with slope $R_B$ across the full observed colour range.
Fig.~\ref{fig:ext_ztf}, obtained from our analysis of the ZTF~HQ~VL sample, shows precisely this latter case: the dust component contributes roughly four times as much to the colour spread as the intrinsic scatter, and a single straight line with slope $R_B$ provides an adequate description of the latent distribution.

\subsection{Comparison with Foundation DR1}\label{subsec:foundation_comparison}

To rule out the possibility that our results are an artefact of the ZTF sample, we re-examine the Foundation DR1 dataset \citep{Foley_2018, Jones_2019}, which spans a similar redshift range as ZTF and has a high degree of homogeneity, making it a natural comparison.

\begin{figure}
    \centering
    \includegraphics[width=1\linewidth]{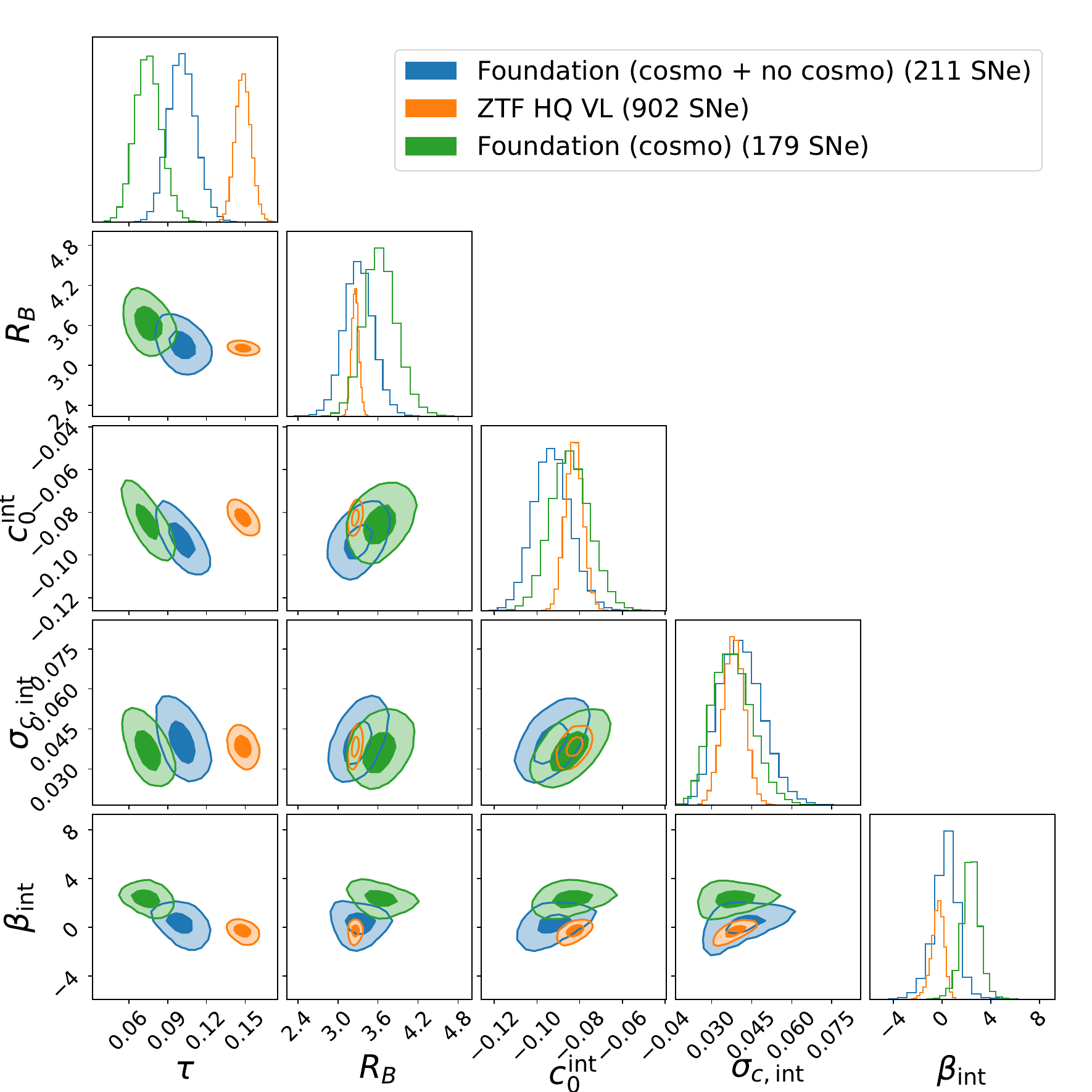}
    \caption{Posteriors of colour-related parameters as fit by Simple--BayeSN for ZTF~HQ~VL and the complete Foundation sample (``cosmo + no cosmo''), compared with results from the ``cosmo'' Foundation sample which discards highly-reddened SNe. Adding the previously excluded redder objects to the Foundation sample improves compatibility with the ZTF results.}
    \label{fig:ztf_fnd_color_params_cornerplot}
\end{figure}

Figure~\ref{fig:ztf_fnd_color_params_cornerplot} compares the inference results from Foundation (with and without the ``no cosmo'' sample) and ZTF. As anticipated from figure~\ref{fig:ztf_foundation_mb_c}, Foundation's ``cosmo'' subset (which has implemented the colour cut $|\hat{c}|<0.3$) implies small amounts of dust reddening ($\tau \approx 0.075$), compensated by a bias in $\beta_{\rm int}$ towards $R_B$. Undoing the colour cut by including the ``no cosmo'' objects shifts the results towards higher $\tau\approx 0.101$ (but still short of ZTF's $\tau\approx 0.148$) and a $\beta_{\rm int}$ consistent with zero. This indicates that our conclusions are not merely ZTF-specific. 
Still, more sophisticated modelling of Foundation's observational selection function would be needed to fully account for the remaining discrepancy, as the domain shift visible in Fig.~\ref{fig:ztf_foundation_mb_c} suggests that survey-level differences between ZTF and Foundation extend beyond the colour cut alone.

\section{Discussion}\label{sec:discussion}

\subsection{Implications for SN standardisation}\label{subsec:discussion}

The results presented in this work bear on the physical interpretation of the Tripp standardisation relation and its nuisance parameters. In standard cosmological analyses, the observed colour--magnitude relation is compressed into a single effective coefficient $\beta_{\rm app}$, assumed to capture both intrinsic SN~Ia colour variations and host-galaxy dust extinction. Our decomposition within Simple-BayeSN challenges the physical interpretation of this parameter, while simultaneously clarifying why the Tripp framework remains empirically effective.

As shown in Fig.~\ref{fig:ext_ztf}, the near-linear geometry of the ZTF~HQ~VL latent colour--magnitude distribution -- in contrast to the curved banana shape observed in M17 -- reflects the dust-dominated regime $\tau \gg \sigma_{c,{\rm int}}$ of the ZTF sample (rather than an intrinsic colour--magnitude correlation) and implies that the linear Tripp relation does not incur the bias that M17 warn of. The Tripp formula is therefore still applicable, despite what M17's parameter estimates suggest, precisely because, as we find, the SN~Ia population is closer to the dust-dominated regime than previously thought.

This is also confirmed empirically by \cite{Ginolin_2025_colour}, who find no statistically significant variation in $\beta_{\rm app}$ when splitting the same ZTF dataset by colour and fitting separate slopes to the blue ($\hat{c}<0$) and red ($\hat{c}>0$) subpopulations: they find $\Delta\beta = -0.71 \pm 0.26$, consistent with zero at a $2.7\sigma$ significance. Our findings also resolve an apparent tension \citet{Ginolin_2025_colour} note: despite clear evidence that blue and red SNe probe physically distinct regimes of the colour distribution, the colour--magnitude relation shows no corresponding inflection.
Within our framework this is not a tension, but the expected outcome: when $\tau \gg \sigma_{\rm c, int}$, the dust component dominates the colour spread across the full observed range, and both subpopulations are governed by approximately the same slope $R_B$, so no change in slope is expected regardless of whether a given supernova's colour originates from intrinsic scatter or dust reddening.

\subsection{Environmental dependences}

\begin{figure*}
    \centering
    \includegraphics[width=0.8\linewidth]{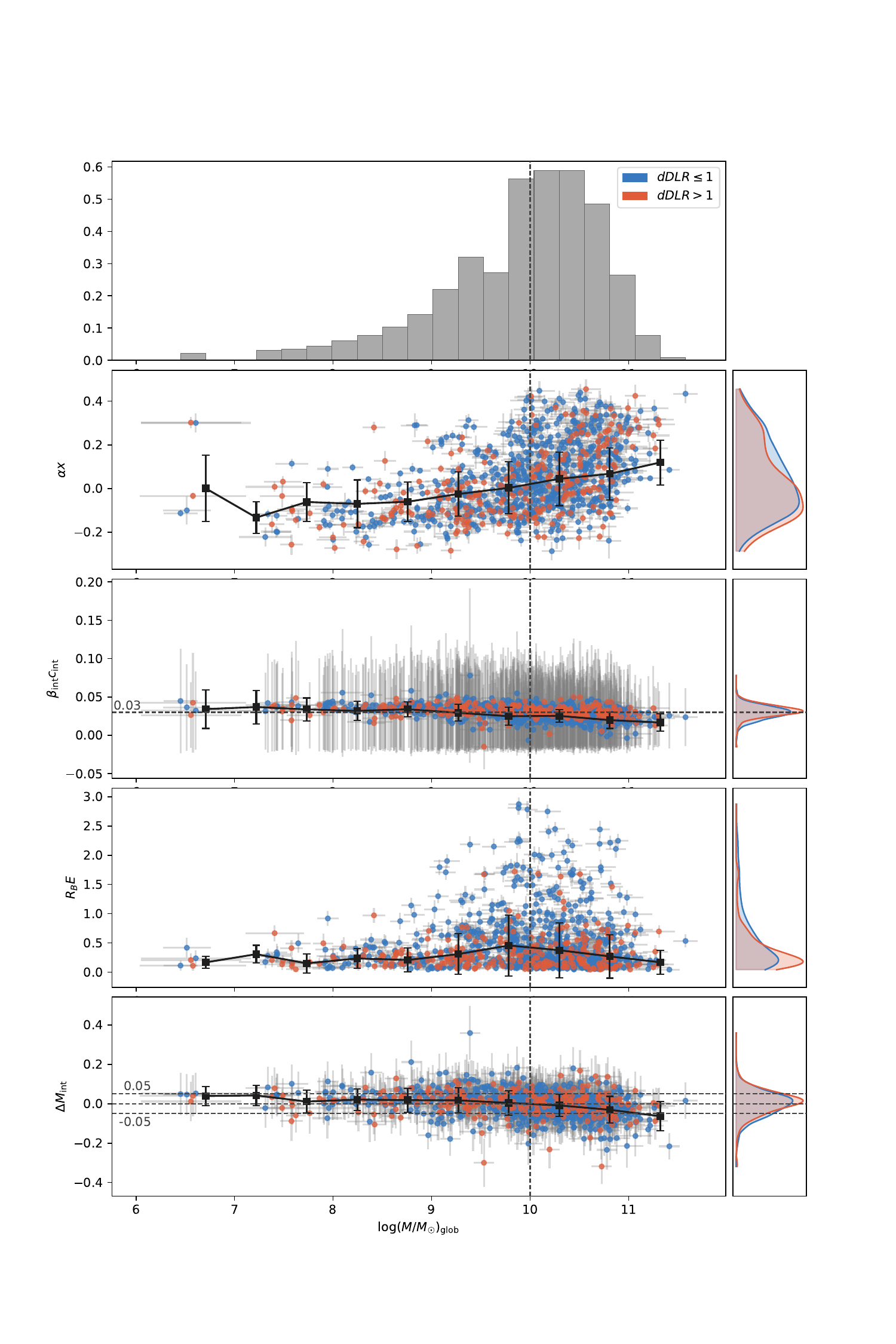}
    \caption{Per-supernova latent contributions to standardised magnitude as a function of host-galaxy stellar mass and coloured according to directional light radius distance ($d_{\rm DLR}$) in the ZTF~HQ~VL sample. Each point corresponds to a SN~Ia, with latent parameters inferred from the posterior mean of the corresponding part of the Gibbs sampler Markov chain. Below the empirical density of the corresponding host galaxy property, the panels show the following contributions to magnitude from top to bottom: effective stretch $\alpha x_s$, effective intrinsic colour $\beta_{\rm int}c_s^{\rm int}$, effective dust extinction $R_B E_s$, and finally the intrinsic magnitude residuals.
    The effective intrinsic colour distribution is consistent with a constant offset across environments, while the dust contribution shows a strong environmental dependence, increasing in higher-mass hosts and at a smaller $d_{\rm DLR}$. A residual correlation between intrinsic magnitude and host mass is visible, consistent with the mass step.}
    \label{fig:ztf_latents_globmass_ddlr}
\end{figure*}
\begin{figure*}
    \centering
    \includegraphics[width=0.8\linewidth]{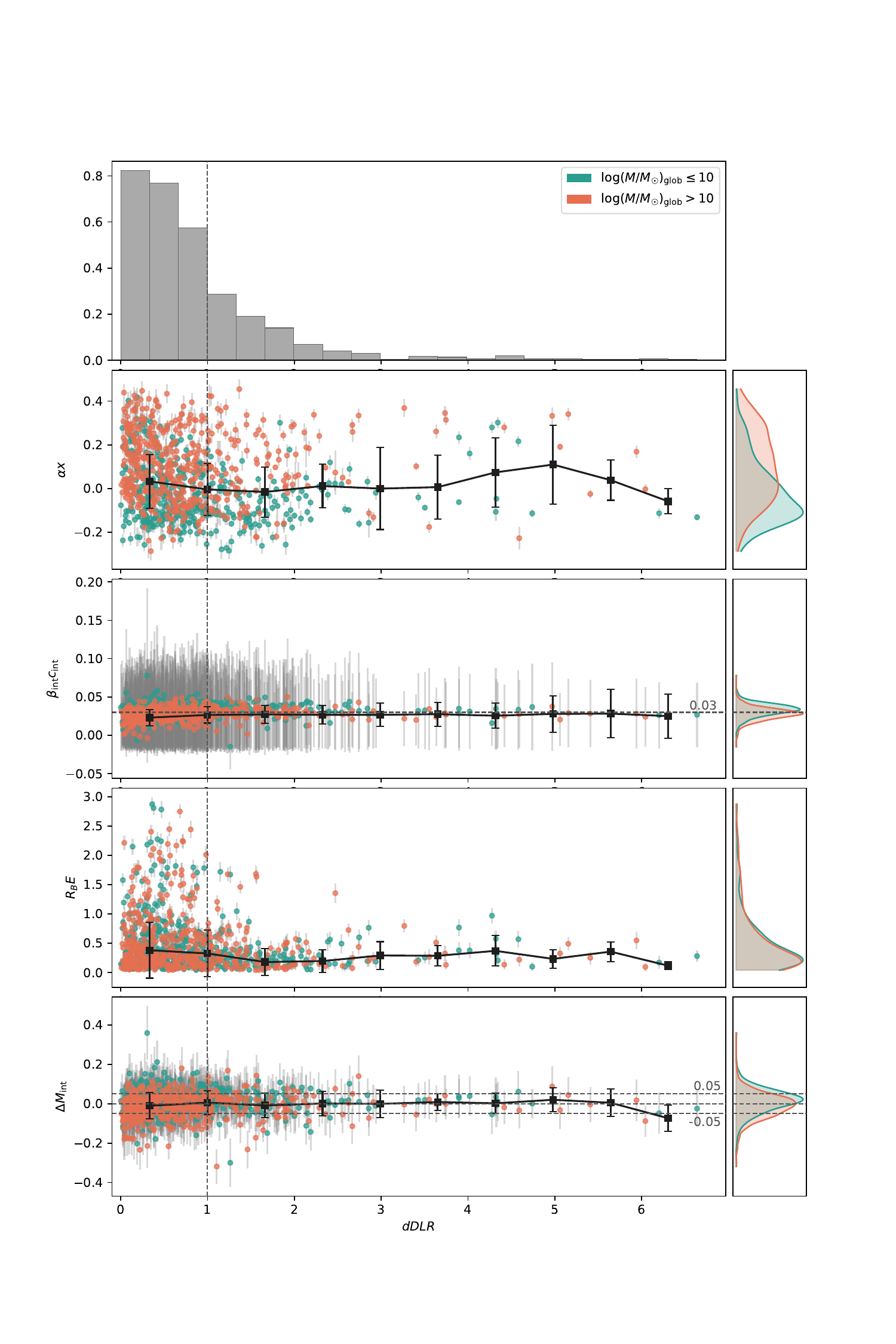}
    \caption{Same as fig.~\ref{fig:ztf_latents_globmass_ddlr} but plotted against $d_{\rm DLR}$ and coloured according to host stellar mass, to highlight complementary projections of the latent parameter space. This representation emphasises the consistency of the inferred trends: dust extinction increases towards smaller $d_{\rm DLR}$ and higher host mass, while effective intrinsic colour contributions remain approximately constant.}
    \label{fig:ztf_latents_ddlr_globmass}
\end{figure*}

\begin{figure*}
    \centering
    \includegraphics[width=1\linewidth]{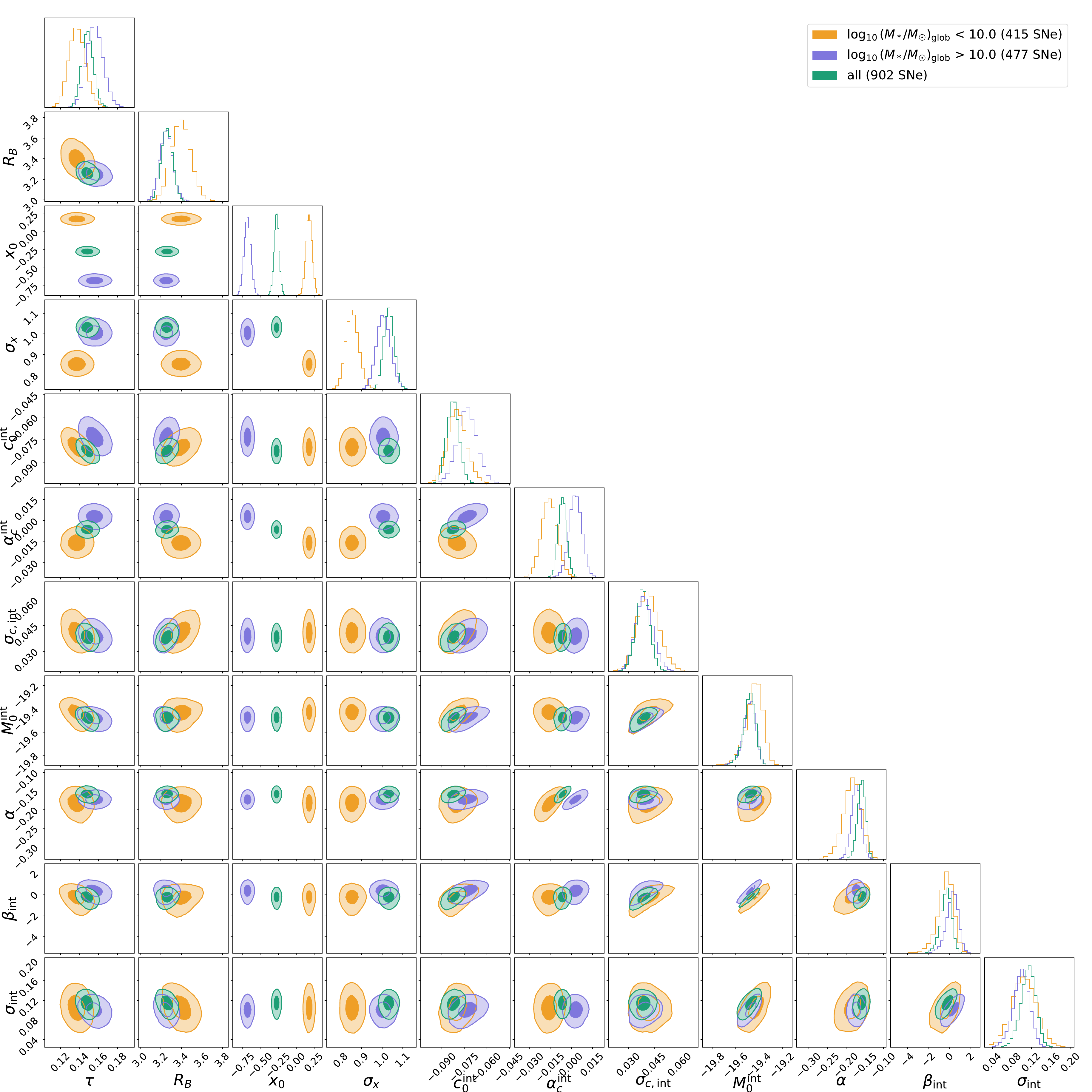}
    \caption{Posterior distributions of Simple-BayeSN population parameters inferred from the subsamples of the ZTF~HQ~VL dataset split by host galaxy stellar mass $\log_{10}(M_*/M_\odot)$. While intrinsic population parameters remain largely unchanged between low- and high-mass hosts, the dust scale $\tau$ increases with host mass -- consistently with the expectation that extrinsic and intrinsic parameters, respectively, should and should not depend on host properties.
    A shift of order $\sim 0.05$~mag in standard intrinsic magnitude parameter $M_0^{\rm int}$ is also observed, corresponding to the SN~Ia mass step. Finally, a well-defined monotonic trend for $x_0$ emerges, which hints at the two-component stretch population previously discussed in the literature \citep[e.g.][]{Ginolin_2025_stretch}.}
    \label{fig:ztf_global_mass}
\end{figure*}
\begin{figure*}
    \centering
    \includegraphics[width=1\linewidth]{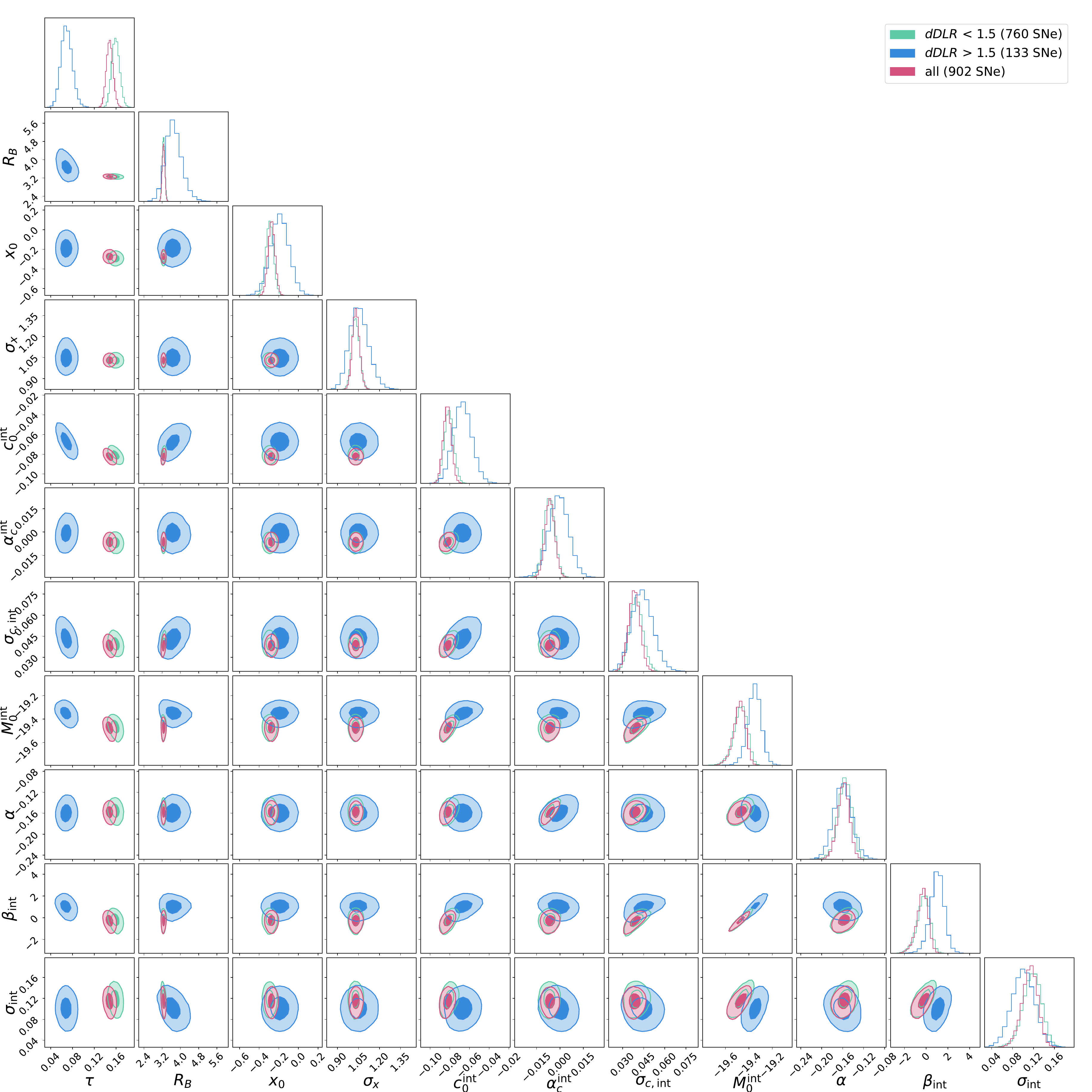}
    \caption{Posterior distribution of Simple-BayeSN population parameters inferred from subsamples of the ZTF~HQ~VL dataset split by directional light radius $d_{\rm DLR}$. The comparison highlights the stability of intrinsic parameters across environments, in contrast with mean dust reddening $\tau$, which scales inversely to $d_{\rm DLR}$, consistent with higher dust column densities in the inner regions of host galaxies.}
    \label{fig:ztf_global_d_dlr}
\end{figure*}

A practical advantage of hierarchical modelling over conventional standardisation is its ability to explicitly separate the various intrinsic and extrinsic contributions ($\alpha x_s$, $\beta_{\rm int} c^{\rm int}_{s}$, $R_B E_s$, and random intrinsic scatter $\Delta M_{\rm int}$) to the overall brightness offset $M^{\rm ext}_s - M_0^{\rm int}$, thus providing a richer vocabulary for studying environmental dependencies. We plot these terms (means and standard deviations a posteriori) versus global host stellar mass and directional light radius distance ($d_{\rm DLR}$) in Figs.~\ref{fig:ztf_latents_globmass_ddlr} and \ref{fig:ztf_latents_ddlr_globmass}.

We clearly observe the well-known trend in SN stretches, discussed by \cite{Ginolin_2025_stretch}, whereby more massive galaxies ($M\gtrsim10^{10} M_\odot$) contain an additional population of high-stretch SNe, and this increases the average stretch-related correction ($\alpha x_s$) in those hosts. We do not investigate this further and note that this effect is still fully captured in our hierarchical modelling (at the per-SN paremeter level) even though we assume a single stretch population rather than a mixture model (and do not model the environment explicitly). Still, by analysing subsamples of SNe hosted in low/high-mass hosts (Fig.~\ref{fig:ztf_global_mass}), we can clearly identify the phenomenon in the difference of posteriors for $x_0$ and $\sigma_x$, which reflect the single low-stretch population for SNe in low-mass hosts and predominantly the high-stretch population otherwise.

The effective intrinsic colour contribution ($\beta_{\rm int} c^{\rm int}_{s}$) is consistent with a constant offset across all environments probed by host-galaxy stellar mass and DLR, reinforcing the interpretation that intrinsic colour variations are small and independent of environment. By contrast, the dust contribution $R_B E_s$ shows a strong and systematic environmental dependence, increasing in higher-mass galaxies and in regions with smaller $d_{\rm DLR}$, consistent with the well-known correlations between stellar mass, galaxy morphology, and dust content.

Another trend we observe concerns the so-called ``mass step'': a difference in the intrinsic scatters $\Delta M_{\rm int}$ (after standardisation and accounting for dust) between SNe in low- and high-mass hosts. A trend on the order $0.05$ mag is noticeable in Fig.~\ref{fig:ztf_latents_globmass_ddlr}, as well as a difference in the standard magnitude $M_0^{\rm int}$ when analysing separately low- and high-mass subsamples (Fig.~\ref{fig:ztf_global_mass}). Importantly, this shift persists after explicitly modelling the dust components with Simple-BayeSN: notice in Fig.~\ref{fig:ztf_global_mass} the higher inferred $\tau$ for high-mass hosts (and, incidentally, for SNe at lower host-centric distances in Fig.~\ref{fig:ztf_global_d_dlr}), indicating that the mass step is not absorbed by the different dust reddening between the two host-mass bins.

As discussed in the introduction, our analysis is not targeted at a definitive characterisation of the mass step: we fix $R_B$ as a single global parameters, whereas a per-SN $R_B$ population with separate dust laws in high- and low- mass hosts -- as advocated by \cite{Brout_2021} and \cite{Popovic_2023} -- could in principle eradicate (part of) this residual. We therefore report the host-mass correlated residual without claiming a qualitative statement about its origin. Detailed hierarchical analyses of the ZTF mass step using more flexible dust models are presented in \cite{Ginolin_2026_bayesn}.

\subsection{Comparison with previous work}

Our results bear most directly on the analyses of \cite{Brout_2021} and \cite{Popovic_2023}, who applied the Gaussian--exponential colour decomposition -- but allowing per-SN values of $\beta_{\rm int}$ and $R_B$, sampled respectively from normal and truncated normal priors -- to a heterogeneous compilation. Several of our results align with those of \cite{Brout_2021} and \cite{Popovic_2023}: our recovered intrinsic colour mean, standard deviation, and dust parameters ($c_0^{\rm int}=-0.083\pm0.004$, $\sigma_{\rm c,int}=0.038\pm0.004$, $R_B=3.260\pm0.055$, $\tau=0.148\pm0.006$), match the values reported in BS21 for their low-redshift compilation with no host-mass split almost exactly ($c_0^{\rm int}=-0.084\pm 0.004$, $\sigma_{\rm c,int}=0.042\pm0.002$, $R_B=3.0\pm0.2$, $\tau=0.17\pm0.04$). Independent confirmation of our intrinsic colour distribution comes from \cite{Ginolin_2025_colour}, who fit a Gaussian--exponential convolution directly to the apparent colour distribution itself, using a very similar sample as our ZTF~HQ~VL dataset -- and recovering values in good agreement with ours, as shown in Tab.~\ref{tab:ztf_mean_std}.

The main difference lies in the intrinsic colour--luminosity slope: 
whereas \cite{Brout_2021} and \cite{Popovic_2023} find $\bar{\beta}_{\rm int} \approx 2$, we recover $\beta_{\rm int}$ consistent with zero. The comparison is not strictly like-for-like: their model fits a \emph{distribution} of $\beta_{\rm int}$ values, whereas our analysis (following M17) fixes a single $\beta_{\rm int}$ across the population. A version of Simple-BayeSN with per-SN $\beta_{\rm int}$ drawn from a population distribution would allow a more direct comparison, and we leave this extension to future work. Even allowing for this, however, their inferred $\bar{\beta}_{\rm int}$ is incompatible with zero at high significance, which may be a consequence  of the heterogeneous composition of the samples used by \cite{Brout_2021} and \cite{Popovic_2023} and the use of an explicit (like in \cite{Popovic_2023}) or implicit (when adopting the Foundation sample) $\hat{c} \leq 0.3$ cut. 
We note, however, that despite this cut \cite{Brout_2021} and \cite{Popovic_2023} nonetheless recover values of $\tau$ broadly comparable with ours.

A complementary approach to ours consists of applying the full BayeSN model \citep{Mandel_2021}, which performs hierarchial Bayesian inference at the light-curve level rather than on fitted summary parameters.
Recently, \cite{Ginolin_2026_bayesn} applied the full BayeSN model to a ZTF sample very similar to ours, with a primary focus on the host mass step rather than on the intrinsic--extrinsic decomposition we study. A direct comparison to our results is, however, difficult, since BayeSN lacks a direct equivalent to SALT's colour parameter and to the standardisation relation (i.e.\@ an explicit correlation between stretch/colour and peak brightness), deriving instead flexible spectro-temporal surfaces for the primary mode of variation and the correlated residual fluctuations.

Using both simulations and a compiled dataset drawn from CSP, CFA~3--4, Foundation, PS1, SDSS, SNLS and DES3YR, BS21 demonstrated that incorrect modelling of the SN~Ia colour population can bias the inferred dark-energy equation of state parameter at the level of $\abs{\Delta w}\approx 0.04$ in the bias correction phase of conventional cosmological analyses.
\cite{Popovic_2023} revisited this with the Pantheon+ sample and found a smaller, but still nonzero, upper-bound systematic of $\abs{\Delta w}\approx 0.014$.
Our revised population parameters would in principle affect distance bias-correction procedures similarly; a quantitative assessment requires forward-modelling analogous to that of \cite{Popovic_2023}, which is beyond the scope of this work.
We note, however, that unlike in M17, the empirical robustness of the linear Tripp relation -- discussed in Sec.~\ref{subsec:discussion} and confirmed by the $\Delta \beta \approx 0$ result in \cite{Ginolin_2025_colour} -- is not compromised by our findings: in the dust-dominated regime we recover ($\tau\gg \sigma_{\rm c,int}$), the near-linear geometry of the colour--magnitude relation ensures that a single effective $\beta_{\rm app}$ provides an adequate description across the full observed colour range.

\section{Conclusions}\label{sec:conclusions}

The present work reassessed the relative importance of intrinsic colour variation and dust-induced reddening and extinction, using a Bayesian hieararchical model applied to the high-quality volume-limited SN~Ia data from ZTF DR2. We showed that the underlying intrinsic--extrinsic decomposition is far more asymmetric than M17, closer to the BS21 regime: $\beta_{\rm int}$ is consistent with zero, and $\sigma_{\rm c, int} \ll \tau$, so that the dust component dominates the colour--magnitude relation. We demonstrated the mechanism explicitly on simulated data and through comparison with the Foundation DR1 sample. A direct consequence is that the colour-tail biases predicted by M17 are largely absent in this dust-dominated regime: the linear Tripp regression provides an adequate description of the colour--magnitude relation across the entire observed colour range, naturally explaining the $\Delta\beta \approx 0$ result of \cite{Ginolin_2025_colour}.

Two features of our analysis are key to this conclusion. Firstly, we work with the full Simple--BayeSN hierarchical model, providing a transparent and statistically rigorous inference of the population parameters that avoids methodological compromises required when working on SALT-based summaries with hand-tuned selection corrections. Secondly, we exploit a homogeneous, volume-limited sample that was not available to M17, BS21, nor \cite{Popovic_2023}, which dispenses with systematics induced by calibration and survey-dependent selection across a heterogeneous compilation.

In the near future, the Vera Rubin Observatory will deliver much larger SN~Ia samples that can be used to further test our findings. In this context, it will be all the more important to refine the modelling, for example by including per-SN variability of the dust law and testing different distributional assumptions, like the Weibull or exponentiated--exponential distributions advocated on the basis of recent simulations by~\cite{Duarte_2026}. An ideal framework for this kind of study is offered by the flexible capabilities of simulation-based inference \citep{Karchev_2022} and particularly the recently introduced joint analysis of SN~Ia and host-galaxy data \citep{Karchev_2026}, which we plan to deploy on ZTF data in the near future.

\section*{Acknowledgements}

The authors thank Kaisey Mandel, Madeleine Ginolin, Benjamin Boyd, Matthew Grayling for useful discussions. 

RT acknowledges co-funding from Next Generation EU, in the context of the National Recovery and Resilience Plan, Investment PE1 – Project FAIR ``Future Artificial Intelligence Research''. This resource was co-financed by the Next Generation EU [DM 1555 del 11.10.22]. RT is partially supported by the Fondazione ICSC, Spoke 3 ``Astrophysics and Cosmos Observations'', Piano Nazionale di Ripresa e Resilienza Project ID CN00000013 ``Italian Research Center on High-Performance Computing, Big Data and Quantum Computing'' funded by MUR Missione 4 Componente 2 Investimento 1.4: Potenziamento strutture di ricerca e creazione di ``campioni nazionali di R\&S (M4C2-19)'' - Next Generation EU (NGEU).

%%%%%%%%%%%%%%%%%%%%%%%%%%%%%%%%%%%%%%%%%%%%%%%%%%
\section*{Data Availability}

\begin{itemize}
    \item The ZTF DR2 sample \citep{Rigault_2025} is publicly available at \url{https://ztfcosmo.in2p3.fr/}.
    \item The Foundation DR1 sample \citep{Foley_2018, Jones_2019} is publicly available at \url{https://github.com/djones1040/Foundation_DR1}.
    \item Our Python implementation of the Simple--BayeSN model, as well as the scripts used to produce the plots in this paper, are publicly available at \url{https://github.com/marco-giunta/ztf-dust}.
\end{itemize}



\bibliographystyle{mnras}
\bibliography{colours_bib}

%%%%%%%%%%%%%%%%%%%%%%%%%%%%%%%%%%%%%%%%%%%%%%%%%%

%%%%%%%%%%%%%%%%% APPENDICES %%%%%%%%%%%%%%%%%%%%%

\appendix

\section{Posterior cornerplots}
\begin{figure*}
    \centering
    \includegraphics[width=1\linewidth]{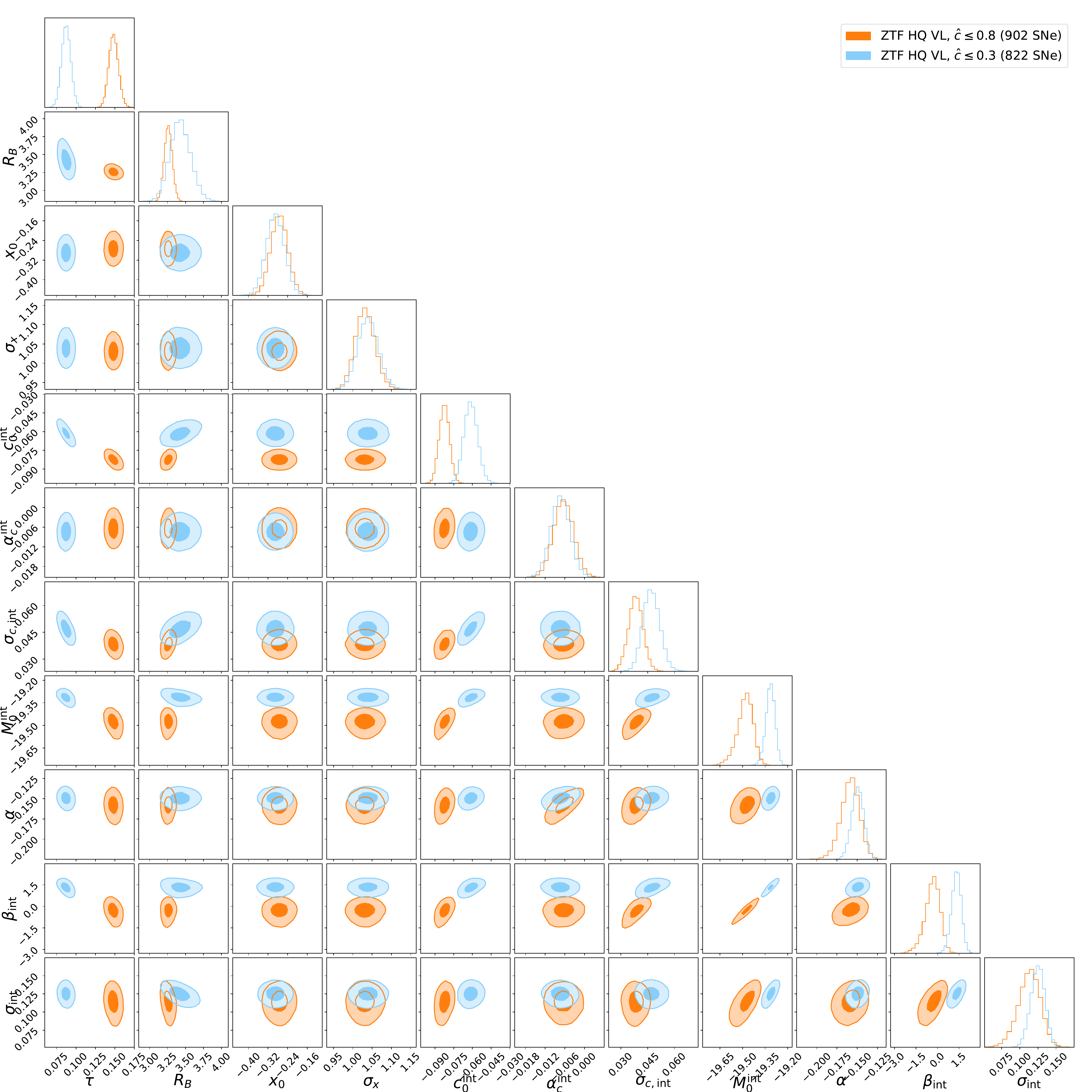}
    \caption{Posteriors for all the Simple--BayeSN parameters fit on the ZTF~HQ~VL sample.}
    \label{fig:ztf_cornerplot}
\end{figure*}

\begin{figure*}
    \centering
    \includegraphics[width=1\linewidth]{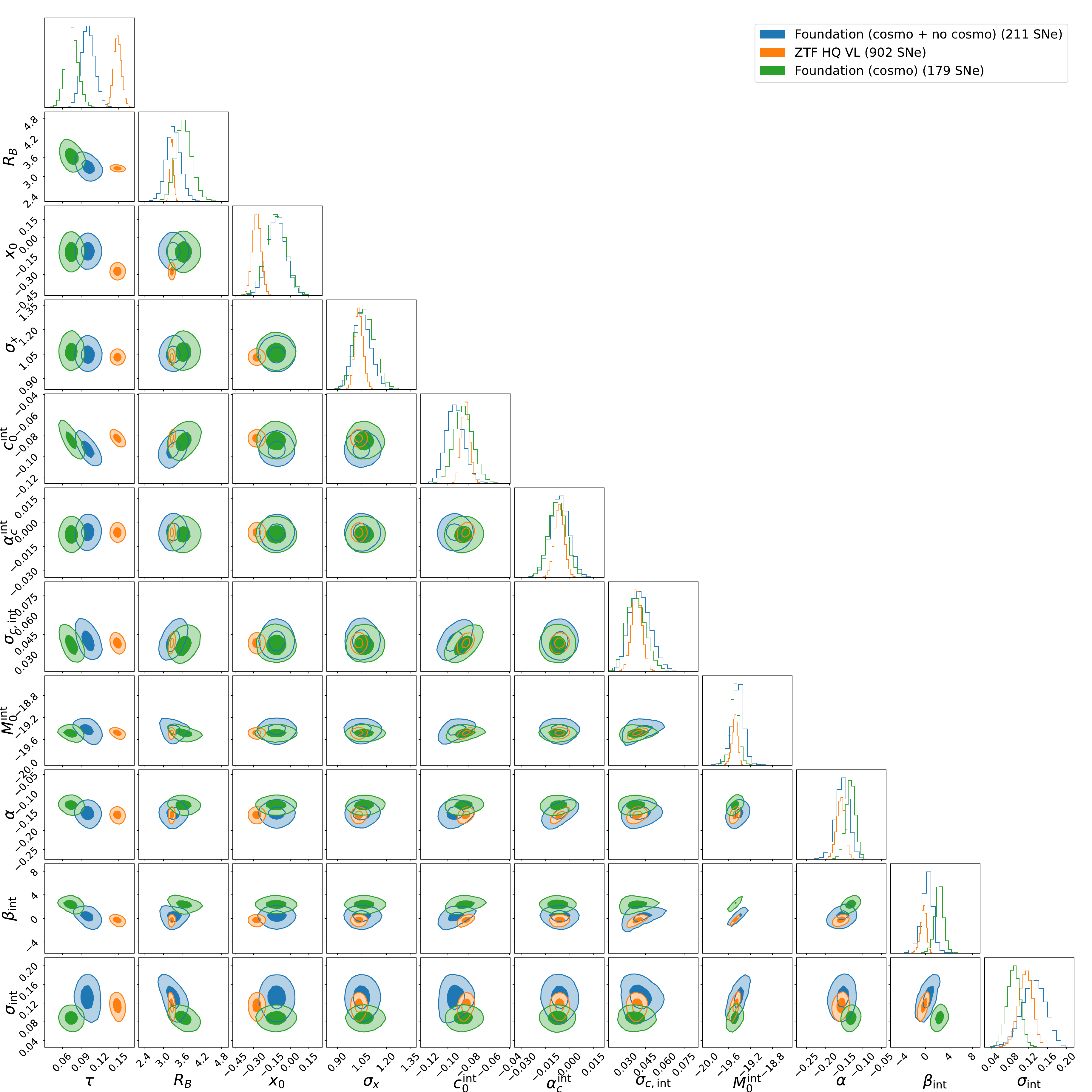}
    \caption{Posteriors for all the Simple--BayeSN parameters fit on the ZTF~HQ~VL sample, and the Foundation sample (with and without the ``no cosmo'' data).}
    \label{fig:ztf_fnd_cornerplot}
\end{figure*}

Figures~\ref{fig:ztf_cornerplot} and \ref{fig:ztf_fnd_cornerplot} show the full posterior distributions for all 11 Simple--BayeSN population parameters.
Figure~\ref{fig:ztf_cornerplot} corresponds to inference on the ZTF~HQ~VL sample alone, with and without the $\abs{\hat{c}}\leq 0.3$ colour cut.
Figure~\ref{fig:ztf_fnd_cornerplot} compares the ZTF~HQ~VL posteriors with those from the Foundation DR1 sample, both with (``cosmo'' only) and without (``cosmo + no cosmo'') the colour cut.

\label{app:cornerplots}
%%%%%%%%%%%%%%%%%%%%%%%%%%%%%%%%%%%%%%%%%%%%%%%%%%

% Don't change these lines
\bsp	% typesetting comment
\label{lastpage}
\end{document}